\documentclass[aps,prd,nofootinbib,preprintnumbers,superscriptaddress,floatfix]{revtex4}

\usepackage{epsfig}
\usepackage{bm}
\usepackage{amssymb}
\usepackage{amsmath}                
\usepackage{color}
\usepackage{subfigure}
\usepackage{dcolumn}

\usepackage[utf8]{inputenc}

\usepackage{amsmath}
\usepackage{amssymb}
\usepackage{amsthm}
\usepackage{amscd, amsfonts, mathrsfs}
\usepackage{cases}
\usepackage{verbatim} 
\usepackage{epsf}
\usepackage[bookmarksnumbered,pdfpagelabels=true,plainpages=false,colorlinks=true,
            linkcolor=black,citecolor=black,urlcolor=black]{hyperref}

\theoremstyle{plain}

\theoremstyle{definition}

\allowdisplaybreaks

\newcommand{\MA}{\mathcal{A}}

\newcommand{\bml}{\begin{subequations}}
\newcommand{\eml}{\end{subequations}}

\newcommand{\be}{\begin{equation}}
\newcommand{\ee}{\end{equation}}
\newcommand{\bea}{\begin{eqnarray}}
\newcommand{\eea}{\end{eqnarray}}

\newcommand{\bbm}{\begin{bmatrix}}
\newcommand{\ebm}{\end{bmatrix}}
\newcommand{\bvm}{\begin{vmatrix}}
\newcommand{\evm}{\end{vmatrix}}

\begin{document}

%%%%%%%%%%%%%%%%%

\title{Cosmological consequences of first-order
general-relativistic viscous fluid dynamics}
\date{\today}

%Authors in alphabetical order

\author{F\'abio S.\ Bemfica}
\affiliation{Escola de Ciências e Tecnologia, Universidade Federal do Rio Grande do Norte, 59072-970, Natal, RN, Brazil}
\email{fabio.bemfica@ect.ufrn.br}

\author{Marcelo M.\ Disconzi}
\affiliation{Department of Mathematics, Vanderbilt University, Nashville, TN 37211, USA}
\email{marcelo.disconzi@vanderbilt.edu}

\author{Jorge Noronha}
\affiliation{Illinois Center for Advanced Studies of the Universe \& Department of Physics, 
University of Illinois at Urbana-Champaign, Urbana, IL 61801-3003, USA}
\email{jn0508@illinois.edu}

\author{Robert J.\ Scherrer}
\affiliation{Department of Physics \& Astronomy, Vanderbilt University, Nashville, TN 37235, USA}
\email{robert.scherrer@vanderbilt.edu}
%\date{\today}
%\pagestyle{empty}

\begin{abstract}
We investigate the out-of-equilibrium dynamics of viscous fluids in a spatially flat Friedmann-Lema\^itre-Robertson-Walker cosmology using
the most general causal and stable viscous energy-momentum tensor defined at first order in spacetime derivatives. In this new framework a  pressureless viscous fluid having equilibrium energy density
$\rho$ can evolve to an asymptotic
future solution in which the Hubble parameter
approaches a constant while $\rho \rightarrow 0$, even in the absence of a cosmological constant (i.e., $\Lambda = 0$). Thus, while viscous effects in
this model
drive an accelerated expansion of the universe, the equilibrium energy density itself vanishes, leaving
behind only the acceleration. This behavior emerges as a consequence of causality in first-order theories of relativistic fluid dynamics and it is fully consistent with Einstein's equations. 
%For completeness, we show that for an appropriate choice of model parameters, a viscous universe in this approach containing only baryons and dark matter (with the latter having nonzero viscosity) can nearly mimic $\Lambda$CDM.

\end{abstract}

\maketitle

%%%%%%%%%%%%%%% INTRODUCTION %%%%%%%%%%%%%%%%%%%%%%%

\section{Introduction}

Given the ubiquity of viscous phenomena around us, it is only natural to wonder how cosmological observations can constrain the presence, or not, of viscous effects at different stages in the large scale evolution of the universe. In fact, dissipative processes in the early universe have been studied for quite some time
\cite{WeinbergViscosityCosmology,Belinsky:1979,Gron:1990ew}.  In isotropic and
homogeneous spacetimes dissipation can only appear from scalar sources,
which motivated early on the study of bulk viscosity in the expansion of
the universe \cite{Murphy:1973zz,Belinsky:1977}. Since this effect is expected to contribute
negatively to the pressure of an expanding universe, bulk viscosity-driven
inflation has also been examined
\cite{Padmanabhan:1987dg,Hiscock:1991sp,Pavon:1990qf,Zakari:1993yhk,Maartens:1995wt,Zimdahl:1996ka}.
More recently, after the discovery of the
current accelerated expansion of the universe \cite{Peebles:2002gy},
the possibility of unifying dark matter and dark energy as a single
viscous fluid generated a lot of
interest \cite{Fabris:2005ts,Colistete:2007xi,LiBarrow,Avelino,Hipolito1,Hipolito2,Gagnon,Piattella_et_al,Velten:2012uv,VeltenSchwarz,BG,Disconzi_Kephart_Scherrer_2015,DisconziKephartScherrerNew,
Cruz:2018psw,Cruz:2019uya} (for a recent review, see \cite{Brevik:2017msy}). 

Before statements concerning the suitability of viscous dark matter models to match cosmological observations can be reliably made, it is important to keep in mind that there are still fundamental theoretical questions concerning the description of viscous effects in general-relativistic fluids that are very relevant to this problem. In fact, in standard approaches viscous processes fundamentally alter the equations of motion of relativistic fluids through the addition of new terms containing space-like derivatives\footnote{Hence, only spatial derivatives appear in the energy-momentum tensor in the fluid's local rest frame.} of the hydrodynamic variables in the fluid's energy-momentum tensor. This occurs, for instance, in the famous theories pioneered by Eckart \cite{EckartViscous} and Landau and Lifshitz \cite{LandauLifshitzFluids}. 
However, these modifications seem to be incompatible with general relativity. Indeed, the theories proposed by Eckart and Landau and Lifshitz are acausal \cite{PichonViscous,Hiscock_Lindblom_instability_1985}, which makes them unsuitable for the investigation of real-time  viscous processes in relativity. Furthermore, it is known that such theories are unstable against perturbations around the thermodynamical equilibrium state  \cite{Hiscock_Lindblom_instability_1985}. This is a consequence of the more general statement that acausal dissipative theories cannot be stable in relativity \cite{Gavassino:2021owo}.

One could at first sight think that the Eckart and Landau and Lifshitz theories
are still suitable for cosmological investigations despite their acausality and
instability because the symmetry assumptions made in cosmological models imply
that the dynamic evolution is described by a system of ordinary differential
equations, whereas causality and stability are concepts applicable only to
partial differential equations.\footnote{One can of course talk about stability of ordinary differential equations. But the type of stability discussed in the context of Eckart and Landau and Lifshitz theories requires at least two independent variables.} Nevertheless, such a description given in terms of ordinary differential equations implicitly assumes that the underlying system of partial differential equations (from which the ordinary differential equations
arise upon imposing symmetry conditions) is well posed. More precisely, since the symmetry assumptions of cosmology are only approximately satisfied (e.g., the universe is not perfectly homogeneous nor isotropic in large scales), the actual system's description is given in terms of the full Einstein-matter equations as partial differential equations. Only if the ordinary differential equations provide a good approximation to the underlying partial differential equations can one take their solutions as a good approximation to the actual, not perfectly symmetric, system. This requires the system of partial differential equations to be well-posed \cite{Rezzolla_Zanotti_book}, a property that fails for the Eckart and Landau and Lifshitz theories \cite{PichonViscous}. 

In sum, the above facts hamper the application of Eckart's and Landau and Lifshitz's theories in  questions concerning the cosmological evolution of the universe. The same can be said about the fate of cosmological fluctuations in such viscous fluid models. Therefore, conclusions obtained from such models must be taken with a grain of salt (at best).

Israel and Stewart (IS) put forward an approach \cite{MIS-5} where linearized disturbances around global equilibrium can be causal and stable \cite{Hiscock_Lindblom_stability_1983}, if certain conditions for the fluid's equation of state and transport coefficients (e.g., bulk and shear viscosities) are fulfilled. However, despite recent progress \cite{Bemfica:2019cop,Bemfica:2020xym}, very little is known about the properties and the constraints that must be fulfilled in these theories in the nonlinear regime, which can be important in simulations already in flat spacetime \cite{Plumberg:2021bme} and, also, when embedding such fluid models in dynamical spacetimes. In fact, in the context of viscous dark fluid modeling, it is not known how the recently found nonlinear constraints \cite{Bemfica:2019cop,Bemfica:2020xym} coming from causality affect previous conclusions drawn from such Israel-Stewart-like models (e.g., \cite{Maartens:1995wt,Cruz:2018psw,Cruz:2019uya}). The constraints become especially relevant in the far-from-equilibrium regime where viscous effects are large, which is probed in viscous dark fluid models that attempt to unify dark matter and dark energy in an accelerating universe.  
In addition, well-posedness of the Israel-Stewart equations remains an open question, except in some very particular cases \cite{Bemfica:2019cop}. This is a potential drawback given the importance
of well-posedness for an accurate description of the evolution, as explained above.

In this work, we investigate how viscous effects can affect the evolution of
fluids in isotropic, homogeneous, and spatially flat spacetimes using the new
general effective-theory formalism originally proposed in
\cite{BemficaDisconziNoronha} and further developed in
\cite{Kovtun:2019hdm,Bemfica:2019knx,Hoult:2020eho,Bemfica:2020zjp}. In this
approach, known as BDNK in the fluid dynamical literature after the initials of
the authors of \cite{BemficaDisconziNoronha,Kovtun:2019hdm}, the viscous
contribution to the energy-momentum tensor is expanded according to a
well-defined power-counting scheme in terms of all the possible time-like and
space-like derivatives of the hydrodynamic variables (e.g. density, flow
velocity) compatible with the symmetries, in contrast to standard formulations 
\cite{EckartViscous,LandauLifshitzFluids} where only terms defined using
space-like derivatives are included. The full system of equations of motion
describing the evolving viscous fluid coupled to Einstein's equations has been
proven  to be causal and strongly hyperbolic \cite{Bemfica:2020zjp}, hence
well-posed, even in the full nonlinear regime.\footnote{Hydrodynamic stability around equilibrium has also been established, see \cite{BemficaDisconziNoronha,Kovtun:2019hdm,Bemfica:2019knx,Hoult:2020eho,Bemfica:2020zjp}.} Therefore, this framework is uniquely suited to investigate real-time dynamical problems, and also mathematical questions, concerning the  coupling of Einstein's equations to viscous fluids in general relativity. Numerical solutions of this theory can already be found in \cite{Pandya:2021ief,Pandya:2022pif,Bantilan:2022ech,Pandya:2022sff} while systematic derivations of BDNK theory from kinetic theory and holography were presented in \cite{Rocha:2022ind,Hoult:2021gnb} (see also \cite{Biswas:2022cla}). Some recent applications can be found  in Refs.\ \cite{Danielsson:2021ykm,Armas:2022wvb}.

We focus in this work on the simple dynamics displayed by a viscous fluid in  Friedmann-Lema\^itre-Robertson-Walker spacetime \cite{Weinberg_GR_book}, in the absence of a cosmological constant ($\Lambda=0$). We show through a variety of examples that an initially dust-like matter component (cold dark matter) can drive
an accelerated expansion at late times when viscous effects are included. This implies that the previous intuition concerning the effects of viscosity acquired from inconsistent, or less well understood, fluid models was well motivated. However, the consistent treatment of causality at all levels in our approach predicts a curious new effect that is only possible in general relativity: the  viscous fluid does not asymptotically achieve
a constant equilibrium density in accelerating cosmologies; instead, this density decays away as a power of the scale factor. This occurs even though the cosmological constant is set to zero. In other words, the equilibrium contribution of the viscous fluid disappears at late times leaving only the acceleration of the universe behind even though there is no cosmological constant. Surprisingly enough, this phenomenon is fully consistent with Einstein's equations. 

This ``Cheshire Cat"-like behavior \cite{carroll:1865} during accelerated
expansion is a consequence of causality in this approach and it cannot be reproduced by any previous model without a cosmological constant where the equilibrium density must remain nonzero when the universe is accelerating. Although different types of behavior for the evolution of the cosmological scale factor are possible, for completeness we show that 
an appropriate choice of model parameters can produce evolution that almost exactly mimics $\Lambda$CDM. While we are not advocating here that viscous fluids provide an alternative way to fully describe cosmological observations by unifying dark energy and dark matter, it is amusing to see that the more sophisticated and theoretically consistent framework employed in this work does not seem to be incompatible with this idea (at least when it comes to the average properties of the universe). 

This paper is organized as follows. In the next section we lay out the general properties of first-order theories of relativistic fluid dynamics and discuss their dynamics in Friedmann-Lema\^itre-Robertson-Walker spacetime. We discuss the case of radiation and also present some new features of such theories in the case of accelerated expansion. In section \ref{fixedpointsection} we analyze the stability property of the equations of motion and their fixed points. In section \ref{viscouscosmo} we consider whether the theory presented here can lead to cosmological evolution consistent
with observations. Our final remarks can be found in section \ref{conclusions}.

\emph{Notation}: We use natural units $\hbar=c=k_B=1$, a 4-dimensional spacetime metric $g_{\mu\nu}$ with a mostly plus signature, and Greek indices
run from 0 to 3 while Latin indices run from 1 to 3. 

\section{Cosmology with first-order general-relativistic viscous fluid dynamics}\label{section2}

Let us briefly review the effective theory approach to relativistic viscous fluid dynamics introduced in Ref.\ \cite{BemficaDisconziNoronha} and further developed in \cite{Kovtun:2019hdm,Bemfica:2019knx,Hoult:2020eho,Bemfica:2020zjp}. As usual \cite{Rezzolla_Zanotti_book}, one starts by decomposing the energy-momentum tensor of a fluid in a general out of equilibrium state in terms of irreducible structures constructed using a time-like future-directed  4-velocity vector $u^\mu$ (where $u_\mu u^\mu=-1$):
\be
T^{\mu\nu}=\mathcal{E}u^\mu u^\nu+ \mathcal{P}\Delta^{\mu\nu}+\pi^{\mu\nu}+u^\mu\mathcal{Q}^\nu+u^\nu\mathcal{Q}^\mu,
\label{general_tensor}
\ee
where $\mathcal{E} = u_\mu u_\nu T^{\mu\nu}$ is the total energy density seen by an observer comoving with the fluid, $\mathcal{P} = \Delta_{\mu\nu}T^{\mu\nu}/3$ is the total fluid pressure defined using the space-like projector $\Delta_{\mu\nu}=g_{\mu\nu}+u_\mu u_\nu$ orthogonal to $u^\mu$, $\mathcal{Q}^\mu = - \Delta^\mu_\lambda u_\nu T^{\nu\lambda}$ describes energy diffusion, and $\pi^{\mu\nu} = \Delta^{\mu\nu\alpha\beta}T_{\alpha\beta}$ is the shear-stress tensor defined using the symmetric, rank-4 traceless projector $\Delta^{\mu\nu\alpha\beta} = \frac{1}{2}\left(\Delta^{\mu\beta}\Delta^{\nu\alpha} + \Delta^{\mu\alpha}\Delta^{\nu\beta}\right)-\frac{1}{3}\Delta^{\mu\nu}\Delta^{\alpha\beta}$ \cite{Rezzolla_Zanotti_book}.

In an equilibrium state, $\pi^{\mu\nu}$ and $\mathcal{Q}^\mu$ vanish, $\mathcal{E}$ becomes the equilibrium energy density $\rho$, and $\mathcal{P}$ the corresponding thermodynamic pressure $P(\rho)$ of the system determined by its equation of state. The corresponding equilibrium energy-momentum tensor is then $T_0^{\mu\nu} = \rho u^\mu u^\nu + P\Delta^{\mu\nu}$. In a general out-of-equilibrium state, one may write $\mathcal{E} = \rho+\mathcal{A}$ and $\mathcal{P} = P + \Pi$, where $\mathcal{A}$ and the bulk scalar $\Pi$ represent the out of equilibrium corrections to the energy density and pressure, respectively, as long as they vanish in equilibrium. In this case, the most general energy-momentum tensor that can describe an out-of-equilibrium state can be written as 
\be
T^{\mu\nu}=(\rho+\MA)u^\mu u^\nu+\left (P(\rho)+\Pi\right )\Delta^{\mu\nu}+\pi^{\mu\nu}+u^\mu\mathcal{Q}^\nu+u^\nu\mathcal{Q}^\mu.
\label{novotensorBDNfodaotche}
\ee
Constraints on the out-of-equilibrium contributions can be readily obtained by the dominant energy condition \cite{WaldBookGR1984}, which imposes that $\rho +\mathcal{A}\geq 0$  and $\mathcal{Q}_\mu\mathcal{Q}^\mu \leq (\rho+\mathcal{A})^2$. This naturally places a bound on the size of some of the out-of-equilibrium corrections. However, we note that the conservation of energy and momentum, $\nabla_\mu T^{\mu\nu}=0$, is not enough to fully determine the evolution described by the 14 dynamical variables $\{\rho,u^\mu, \mathcal{A},\Pi,\mathcal{Q}^\mu, \pi^{\mu\nu}\}$. Therefore, some  procedure must be implemented to fully specify the system's dynamics. Instead of treating the non-equilibrium corrections as new degrees of freedom (and consequently postulating new additional equations of motion for them) as in Israel-Stewart-based approaches and extended irreversible thermodynamics \cite{JouetallBook}, here we consider the case where the effective theory describing the macroscopic motion of the system is defined solely in terms of the standard hydrodynamic variables already present in equilibrium, which in our case are $\{\rho,u^\mu\}$. In this approach,
the dissipative contributions must be given in terms of the hydrodynamic fields $\{\rho,u^\mu\}$ and their derivatives, which may be organized through a relativistic derivative expansion. Assuming that deviations from equilibrium are small, the most general theory compatible with the symmetries that can be written in terms of first-order derivatives is defined by
\be
\MA=\chi_1\frac{u^\alpha\nabla_\alpha \rho}{\rho+P}+\chi_2 \nabla_\alpha u^\alpha,\,\, \Pi=\chi_3\frac{u^\alpha\nabla_\alpha \rho}{\rho+P}+\chi_4 \nabla_\alpha u^\alpha,\,\, 
\mathcal{Q}_\mu=\lambda\left (\frac{\Delta^{\nu}_\mu\nabla_\nu P}{\rho+P}+u^\alpha\nabla_\alpha u_\mu\right), \,\, \pi_{\mu\nu} = -2\eta \sigma_{\mu\nu},
\label{BDNdef}
\ee
where $\sigma_{\mu\nu}=\Delta_{\mu\nu}^{\alpha\beta}\nabla_\alpha u_\beta$ is the shear tensor.
Above, the shear viscosity $\eta$, and the coefficients $\lambda$ and $\chi_1$, $\chi_2$, $\chi_3$, and $\chi_4$ are in principle known functions of $\rho$, which are determined from the underlying microscopic theory. The bulk viscosity coefficient is given by the combination \cite{Kovtun:2019hdm,Bemfica:2019knx} 
\be
\zeta = \chi_3-\chi_4 + c_s^2 (\chi_2-\chi_1),
\ee
where $c_s^2 = dP/d\rho$ is the equilibrium speed of sound squared. The transport coefficients $\eta$ and $\zeta$ determine how the long wavelength limit of hydrodynamic modes (i.e., sound and shear disturbances) are damped and the amount of entropy produced \cite{Kovtun:2019hdm}, while three out of the four $\chi$ coefficients determine the scales associated with non-hydrodynamic\footnote{Those describe linearized disturbances around equilibrium that carry energy even in the homogeneous limit.} modes.     

It is important to stress a few properties of the expressions above. First, given that in equilibrium $u_\mu/T$ (with $T$ being the temperature) is a Killing vector \cite{Hiscock_Lindblom_stability_1983} and $\rho = \rho(T)$, every single term in $\mathcal{A}$ and $\Pi$ separately vanishes in equilibrium, while  $ \frac{\Delta^{\nu}_\mu\nabla_\nu T}{T}+u^\alpha\nabla_\alpha u_\mu=0 \Longrightarrow \mathcal{Q}_\mu=0$. We also note that in this approach time-like derivatives of the density, $u^\alpha \nabla_\alpha \rho$, appear in the constitutive relations. This fact is crucial for ensuring that the evolution is causal \cite{BemficaDisconziNoronha} and linearly stable around equilibrium. Indeed, precise conditions for the coefficients $\{\eta,\lambda,\chi_1,\chi_2,\chi_3,\chi_4\}$ can be found that guarantee causality, stability, strong hyperbolicity and, thus, local well-posedness of solutions of Einstein's equations coupled to the viscous fluid equations \cite{Bemfica:2019knx,Bemfica:2020zjp}. These conditions are violated in the Landau-Lifshitz theory \cite{LandauLifshitzFluids}, which corresponds to setting $\chi_1=\chi_2=\chi_3=\lambda=0$ and $\chi_4 = -\zeta$.

%Finally, we note that the conformal tensor proposed in \cite{BemficaDisconziNoronha} is recovered when $P=\rho/3$ and $\chi_1=\chi_2 = \chi$ and $\chi_3=\chi_4=\chi/3$ in Eq.\ \eqref{BDNdef}.

In this work we initiate the investigation of the possible cosmological consequences of this approach. We consider the viscous fluid theory defined by  \eqref{novotensorBDNfodaotche} and \eqref{BDNdef} coupled to Einstein's equations (without a cosmological constant) in spatially flat Friedmann-Lema\^itre-Robertson-Walker (FLRW) spacetime described by the line element \cite{WeinbergCosmology}
\be
\label{1FRW}
ds^2=-dt^2+a^2(t)\delta_{ij}dx^i dx^j,
\ee
where $a(t)$ is the cosmological scale factor.
Homogeneity and isotropy impose that the shear-stress tensor and the energy diffusion exactly vanish, and our tensor becomes simply 
\be
T^{\mu\nu}=\left(\rho+\chi_1\frac{u^\alpha\nabla_\alpha \rho}{\rho+P}+\chi_2 \nabla_\alpha u^\alpha\right)u^\mu u^\nu+\left (P+\chi_3\frac{u^\alpha\nabla_\alpha \rho}{\rho+P}+\chi_4 \nabla_\alpha u^\alpha\right )\Delta^{\mu\nu}.
\ee
We use $u^\alpha=(1,0,0,0)$ and $\nabla_\mu u^\mu = 3H(t)$, where $H(t)=\dot{a}(t)/a(t)$ is the Hubble expansion rate (with notation $\dot{a}=da/dt$). Using \eqref{1FRW}, Einstein's equations dictate that the spatial derivative $\partial_i\rho=0$, and one finds the following set of equations of motion for $\rho$ and $H$:
\bml
\label{11FRW}
\bea
\dot{H}+H^2&=&-\frac{1}{6}\left [\rho+3P+
\frac{3\chi_3+\chi_1}{\rho+P}\dot{\rho}
+3(3\chi_4+\chi_2)H\right ],\label{11aFRW}\\
H^2&=&\frac{1}{3}\left (\rho+\frac{\chi_1}{\rho+P}\dot{\rho}
+3\chi_2 H\right ),\label{11bFRW}
\eea
\eml 
where we have appropriately normalized the fields above to take into account the $8 \pi G$ constant factor present in Einstein's equations. 
As in the ideal fluid case, the equations of motion of the fluid $\nabla_\mu T^{\mu\nu}=0$ follow directly from those above. Since causality requires $\chi_1>0$ \cite{Bemfica:2019knx}, it is convenient to rewrite \eqref{11aFRW} as
\be
\label{13FRW}
\dot{H}+H^2=-\frac{1}{2}\left [P-\frac{\chi_3}{\chi_1}\rho+\frac{\chi_1+3\chi_3}{\chi_1}H^2-3\chi_2\left (\frac{\chi_3}{\chi_1}-\frac{ \chi_4}{\chi_2}\right )H\right ].
\ee
Finally, it is useful to define the variable 
\be
w=\frac{P}{\rho}
\ee
to investigate the out-of-equilibrium properties of the fluid for different types of equation of state.  

\subsection{Radiation}

For radiation $w=1/3$ and the imposition of  conformal invariance implies that $\chi=\chi_1=\chi_2 = 3\chi_3=3\chi_4\sim \rho^{3/4}$ (hence, $\zeta=0$) \cite{BemficaDisconziNoronha} and we obtain from \eqref{11bFRW} and \eqref{13FRW}
\be
\dot\rho + 3 H \rho \left(1+w_{eff}^{(r)}\right)=0 \qquad \textrm{and} \qquad  \dot{H}+2H^2 = 0,
\label{raditioneq}
\ee
where $w_{eff}^{(r)} = \frac{1}{3} + \frac{4}{9H\chi}\left(\rho-{3H^2}\right)$.
One recognizes that the Hubble parameter decouples from the energy density as it obeys the well-known
equation found for radiation in equilibrium in FLRW \cite{WeinbergCosmology},
with general solution $H(t) = H_0/(1+2H_0 t)$. Furthermore, we note that $\rho = 3H^2$ is a solution of the equation of motion for the energy density. Though at first one may think that the equation of motion for $\rho$ in \eqref{raditioneq}
should have a complicated solution, the uniqueness property to the solutions of the equations of motion (i.e., well-posedness)
directly implies that the general solution for the energy density equation is indeed simply $\rho = 3H^2$, just as in the ideal fluid case. Therefore, as expected, out-of-equilibrium corrections vanish exactly for pure (conformal) radiation in FLRW where $\zeta=0$.
This is also true in the Landau-Lifshitz theory. 

Besides indicating theoretical consistency, this result is also important from the standpoint of observational cosmology as the behavior of the universe when it is dominated by
radiation is tightly constrained both by Big Bang nucleosynthesis (BBN) \cite{Cyburt:2015mya} and by observations of fluctuations in
the cosmic microwave background (CMB) \cite{Planck2018}.  Thus, any model for viscosity that
strongly alters the expansion history of the universe
during the radiation-dominated era can be ruled out.  As our
theory produces no change in the radiation equation of state, it automatically
satisfies this constraint.

\subsection{Zero entropy production limit away from conformal invariance}

Above, we saw that for conformal radiation  viscous effects drop out of the Friedmann equation entirely and no entropy is produced, as expected.  We
have no similar requirement on the values of the $\chi$'s for other equations of state,
but we can derive the requirements on these parameters such that out-of-equilibrium corrections
vanish. One can see that if $\chi_3=\chi_4$ in \eqref{13FRW} and $\chi_1=\chi_2$ in \eqref{11bFRW}, then again $3\rho=H^2$ is a solution of the equation of the equations of motion, which reduce to those of an ideal fluid. Again, well-posedness implies that the solutions are unique and, thus, as long as these two conditions for the $\chi$ parameters are satisfied, the out-of-equilibrium corrections
to the Friedmann equations vanish.  Note that this conclusion is independent of
the $\rho$-dependence of any of the $\chi$ parameters. Also, we remark that when $\chi_1=\chi_2$ and $\chi_3=\chi_4$ one finds that $\zeta=0$, so indeed no entropy \cite{Kovtun:2019hdm} is produced in this case.

\subsection{Landau-Lishitz theory}

In the absence of a cosmological constant, in Landau-Lifshitz theory where $\chi_1=\chi_2=\chi_3=0$ and $\chi_4=-\zeta$  the
viscous dark matter evolves, at late times, to a fluid with a constant density thus mimicking the evolution of dark energy. This can be easily understood because in this model there are no
out-of-equilibrium corrections to the energy density ($\mathcal{A}=0$) and Einstein's equations
become (taking $w=0$ for simplicity)
\be
H^2 = \frac{\rho}{3} \qquad \mathrm{and}\qquad  \dot{H}+H^2 = -\frac{1}{6}\left(\rho - 9 \zeta
H\right).
\ee 
One can see that here a constant energy density implies a constant Hubble expansion rate,
with nonzero solution $H = \zeta$ that is positive for $\zeta>0$. Therefore, in Landau-Lifshitz theory
it is possible for viscous matter to behave at late times as dark energy at the background
level.\footnote{The same result holds for standard IS theories such as those considered in \cite{BemficaDisconziNoronha_IS_bulk}.} This result motivated the creation of many unified viscous dark matter scenarios \cite{Brevik:2017msy} and, in fact, it is known that this model can provide a good description of the background expansion of the universe \cite{LiBarrow}. However, density perturbations in this scenario based on Landau-Lifshitz theory are rapidly damped out, which leads to severe constraints when attempting to
reconcile it with precision cosmology data \cite{LiBarrow,Velten:2011bg,Velten:2012uv}.

We emphasize that such constraints relied on theories known to be acausal
or for which nonlinear causality remains open. We contend that decisive conclusions about the 
viability of the viscous dark matter idea should be based exclusively 
on viscous theories that satisfy causality and well-posedness, in the nonlinear
regime and also when  coupling to Einstein's equations, and for which
linear stability in flat spacetime also holds. In Ref.\ \cite{BemficaDisconziNoronha_IS_bulk} it was proven that IS theories, in the absence of shear viscosity and heat flow, fulfill these requirements (we note that cosmological perturbations in bulk viscous IS-like theories were studied in \cite{Piattella_et_al}). However, we point out that since effects from shear and heat flow do contribute when the spacetime is not homogeneous and isotropic, their influence on the evolution of cosmological perturbations must also be investigated \cite{Barbosa:2017ojt}. To the best of our knowledge, the general first-order theory studied here is now the only framework that fulfills the consistency conditions mentioned above and can, thus, be used to reliably study the effects of bulk, shear, and heat flow even in viscous inhomogeneous cosmological applications in a model independent manner. 
 
\subsection{The Cheshire Cat mechanism} 

We show below that our approach differs from earlier attempts to model dark energy as a viscous phenomenon in an interesting way.  
%When $w=0$ one finds from \eqref{11bFRW} and \eqref{13FRW}  
%\be
%\dot\rho + 3 H \rho \left(1+w_{eff}^{(m)}\right)=0 \qquad \textrm{and} \qquad
%\dot{H}+\frac{3H^2}{2} = -\frac{1}{2}\left[3H \chi_4 - 3 H \chi_3 (1+w_{eff}^{m})\right] 
%\label{eqsmatter}
%\ee
%where\footnote{We note that the corresponding ideal fluid limit can be obtained by multiplying \eqref{eqsmatter} by
%$\chi_1$ and subsequently taking the limit of all transport coefficients equal to zero.} $w_{eff}^{(m)} =-1
%+\frac{\chi_2}{\chi_1} +\frac{1}{3H\chi_1}\left(\rho-{3H^2}\right)$. It is easy to see that constant $\rho$
%implies that $w_{eff}^{(m)} =-1$ and that $H$ also has to be constant. The second equation in \eqref{eqsmatter} then
%implies that $\chi_4 H < 0$ (we note that $H=0$ is excluded by the form of $w_{eff}^{(m)}$). Therefore, the causality
%condition $\chi_4 > 0$ is only preserved for $H < 0$. One then concludes that causality, a core concept of general relativity,
%forbids single component viscous matter to behave at late times as some form of constant dark energy
%in an expanding universe. Other solutions, where $-1<w_{eff}^{(m)}<0$ and $H$ varies sufficiently slowly with time, will be studied %elsewhere. 
In the model investigated here, the equilibrium component of the density of the viscous dark matter is driven asymptotically to zero.
Because of the way that the Friedmann equation is altered by viscous effects in this model,
the universe continues to accelerate even after the driver of this acceleration effectively disappears. Therefore,
we refer to this as the ``Cheshire Cat" mechanism for generating accelerated expansion\footnote{Note that the term ``Cheshire Cat" has been used
in an entirely different way in the context of quantum measurement theory \cite{Aharonov}.}.  

This effect can be most easily illustrated by the following analytical example. For simplicity, let us assume a constant non-negative $w \ll 1$ so the viscous matter has very small equilibrium pressure, which works as
the small parameter in the perturbative argument that follows.
Assume that the matter is such that $\eta\geq 0$ and $\zeta=2\eta w (\frac{1}{3}-w)$. The conditions for causality, well-posedness,
and stability are satisfied if, for instance, $\chi_1=\lambda=4\eta w$, $\chi_2 = 2\eta(1-w)$, $\chi_3 = w \chi_1$, and $\chi_4=4\eta w/3$ \cite{Bemfica:2019knx}.
We note that the viscosity coefficients are, thus, very small and when $w=0$ we recover an ideal fluid
with a matter-like, pressureless equation of state\footnote{Note that
$\eta$ and $\lambda$ need not to be zero even though shear and heat flux vanish. The latter vanish because of symmetries and not because the
coefficients are zero.}.
We take $\eta$ (and, thus, $\zeta$) to be constant. Under these conditions, assuming that $H$ is constant and non-negative, the general solution of the equations of motion \eqref{11bFRW} and \eqref{13FRW} can be found analytically 
\be
H = H_0 =\frac{\zeta}{1+w} \qquad \textrm{and} \qquad \rho(t) = \frac{\rho_0\,\beta}{(\rho_0+\beta)e^{t\beta/\alpha}-\rho_0},
\ee
where $\alpha = 4\eta w/(1+w)$ and $\beta = \alpha H_0(3-w)/2w$. Therefore, even though $H$ is constant,
we note that the energy density still varies with time, decreasing exponentially
towards zero even though there is no cosmological constant\footnote{For a comparison, consider the evolution of an ideal fluid with equation of state given by $P=w\rho$, in the presence
of a positive cosmological constant $\Lambda$ (constant dark energy).
Asymptotically, $3H^2 \to \Lambda$ and the ideal fluid energy density vanishes.
The difference here in the viscous case is that the energy density evolves toward zero
when $H$ is constant and in the absence of a cosmological constant.}. We remark that for this type of matter the Hubble constant is very small, as small as $w$, and that the energy density still varies in time (i.e, it is not a cosmological constant). 

In an accelerated expansion driven by dark energy,
the universe expands at a constant rate and the dark
energy density remains constant.
In the example above, the cosmological scale factor $a(t)$ increases exponentially
but causality generally imposes that the energy density must lag behind, varying on time scales of the order of $\chi_1/(\rho(1+w))$. According to the theory presented here, this is
a general consequence of causality and energy-momentum conservation in out of equilibrium systems described by first-order theories of relativistic viscous fluid dynamics. We shall return to the Cheshire Cat mechanism in Section \ref{viscouscosmo}. \\

\section{Fixed point analysis}\label{fixedpointsection}

Here we set $w=0$, i.e., $P=0$, and study the fixed points and the stability properties of the equations of motion. We also assume that $\chi_1 >0$. We start by rewriting the relevant equations of motion as
\be
\chi_1\dot\rho + 3 H \rho \chi_2  + \rho\left(\rho-{3H^2}\right)=0
\ee
and
\be
2 \chi_1 {\dot{H}} = \chi_3 \rho - {3H^2}\left(\chi_1+\chi_3\right)+3H\left(\chi_2\chi_3-\chi_1\chi_4\right).
\ee
We assume as before that $\chi_i = \chi_i(\rho)$. Let $\rho_0$ and $H_0$ be the fixed points of the equations above so $\dot{H_0}=\dot{\rho_0}=0$. Consider now fluctuations around the fixed points $\rho(t) \to \rho_0 + \delta \rho(t)$ and $H(t)\to H_0 + \delta H(t)$. Note that fluctuations also act on the transport coefficients, i.e., $\chi_i(\rho) \to \chi_i(\rho_0) +  \chi_i'(\rho_0)\delta \rho(t)$, where $\chi_i' = d\chi_i/d\rho$. To zeroth order in the fluctuations we obtain 
\be
 \rho_0\left(\rho_0+3 H_0  \chi_2(\rho_0)  - {3H_0^2}\right)=0
 \label{0thorder_rho}
\ee
and
\be
3H_0 \chi_1(\rho_0)\left({H_0}+\chi_4(\rho_0)\right) =
\chi_3(\rho_0)\left(\rho_0 + 3H_0 \chi_2(\rho_0)- {3H_0^2}\right).
 \label{0thorder_H}
\ee
We see that Eq.\ \eqref{0thorder_rho} implies that $\rho_0=0$ or $\rho_0+3 H_0 
\chi_2(\rho_0)= {3H_0^2}$. Clearly, the latter can be complicated since one must know how $\chi_2$ depends on $\rho$ to solve it. 

\subsection{$\rho_0=0$ and $H_0=0$}

It is easy to see that $\rho_0=H_0=0$ is a fixed point. In fact, this fixed point is very general as it does not depend on the properties of the $\chi_i$'s. Let us now study the linear stability properties of this fixed point. The linearized equations for the fluctuations are
\bea
\delta \dot\rho &=& 0, \\
2 \delta \dot{H} &=& \frac{\chi_3(0)}{\chi_1(0)}\delta \rho + 3\left(\chi_2(0)\chi_3(0)-\chi_1(0)\chi_4(0)\right)\delta H.
\eea
We can write this in matrix form, which reveals that the
eigenvalues of the matrix
are $0$ and $(3/2)\left(\chi_2(0)\chi_3(0)-\chi_1(0)\chi_4(0)\right)$.
Since one of the eigenvalues vanishes
(in fact, the determinant of the matrix vanishes),
this is a marginal case where one does not have an isolated fixed point. In
this case, a linear stability analysis is not guaranteed to give
the correct information about the stability properties of the
system \cite{strogatz:2000}.  In any event, this is not a physically interesting
case.

\subsection{$\rho_0=0$ and $H_0\neq 0$}

In this case, assuming that $\chi_1(0)+\chi_3(0)\neq 0$, one finds
\be
H_0 = \frac{\left(\chi_2(0)\chi_3(0)-\chi_1(0)\chi_4(0)\right)}{\chi_1(0)+\chi_3(0)}.
\ee
We will now find the conditions that ensure that $H_0>0$ and $\rho_0=0$ is an attractor (i.e., a stable isolated fixed point). In this case, the equations for the linearized fluctuations $\vec{v} = {\delta \rho \choose \delta H}$ become 
\be
\frac{d\vec{v}}{dt} = A\, \vec{v},
\ee
where $A$ is a 2 x 2 matrix that depends on $\chi_i(0)$ and $\chi_i'(0)$. The eigenvalues of $A$ are
\be
a_1 = {3H_0}\left(H_0-  \chi_2(0)\right),\qquad a_2 = -\frac{3}{2 \chi_1(0)}\left[\chi_2(0) \chi_3(0)-\chi_1(0)\chi_4(0)\right].
\ee
Thus, a stable fixed point occurs when $a_1 < 0$ and $a_2 <0$ (note that both eigenvalues are real). We see that $a_2 < 0$ implies that  
\be
\chi_2(0) \chi_3(0)-\chi_1(0)\chi_4(0)>0.
\ee
Since that quantity appears in $H_0$, which we consider to be positive, we see that this occurs then if $\chi_1(0)+\chi_3(0)>0$. The condition that $a_1<0$ then occurs when $\chi_4 > -\chi_2$. When those conditions are met, the fixed point is \emph{hyperbolic} (see Chapter 6 of Ref.\ \cite{strogatz:2000}), which means that the qualitative behavior of the system's phase portrait near the attractor is not changed even when small nonlinear terms are included. As a matter of fact, the Hartman-Grobman theorem \cite{strogatz:2000} states that the local phase portrait near a hyperbolic fixed point is topologically equivalent to the phase portrait obtained via linearization and, thus, the conclusions regarding the stability of the fixed point are the same as in the linearized system (in other words, the phase portrait near this attractor is structurally stable). 

Summarizing, we conclude that $\rho=0$ with constant $H_0>0$ is an attractor when the following conditions are met:
\bea
&&\chi_2(0) \chi_3(0)-\chi_1(0)\chi_4(0)>0\\
&& \chi_1(0)+\chi_3(0)>0 \\
&& \chi_4 > -\chi_2,
\eea
which are compatible with the conditions found for causality and also linear stability around equilibrium presented in Ref.\ \cite{Bemfica:2019knx}.

%%%%%%%%%%%%%%% SECTION %%%%%%%%%%%%%%%%%%%%%%%%%%%%

\section{Toward a Realistic Cosmology}\label{viscouscosmo}

In this section we investigate whether the viscous fluid  presented here can
produce cosmological evolution consistent with observations, within a variety of different scenarios.  In particular, we will
be interested in the types of models discussed in Refs.\
\cite{Avelino,Hipolito1} in which the universe contains a pressureless
dark matter component whose viscosity drives the accelerated expansion.

From an observational point of view, it is useful to redefine our
evolution equations in terms of the scale factor $a$ instead of the time.
Measurements of the dark energy density are effectively determinations
of $H(z)$, where the redshift $z$ is related to the scale factor as
$a = 1/(1+z)$.  Using $d/dt = Ha\,d/da$, we can rewrite Eqs.\
(\ref{11aFRW})-(\ref{11bFRW}) with the scale factor as the independent variable:
\begin{eqnarray}
\label{Fried1}
H^\prime a + \frac{3}{2}H &=& -\frac{P}{2H} -\frac{1}{2} \chi_3 \frac{\rho^\prime}{\rho +P} a - \frac{3}{2} \chi_4,\\
\label{Fried2}
H^2 &=& \frac{1}{3} \left( \rho + \chi_1 \frac {\rho^\prime}{\rho + P}aH
+ 3 \chi_2 H \right),
\end{eqnarray}
where we denote
${}^\prime = \frac{d}{da}$. When $P=0$, Eqs.\ (\ref{Fried1}) and (\ref{Fried2}) become
\begin{eqnarray}
\label{Fried1cdm}
H^\prime a + \frac{3}{2}H &=& -\frac{1}{2} \chi_3 \frac{\rho^\prime}{\rho} a - \frac{3}{2} \chi_4,\\
\label{Fried2cdm}
H^2 &=& \frac{1}{3} \left( \rho + \chi_1 \frac {\rho^\prime}{\rho}aH
+ 3 \chi_2 H \right).
\end{eqnarray}
The $\chi$ parameters must satisfy the conditions for nonlinear causality and
linear stability, which were obtained in Ref.\ \cite{Bemfica:2019knx}.  When $P=0$ these conditions are
\begin{eqnarray}
\label{condition1}
\lambda \chi_1 \ge \lambda \chi_3 + \chi_2  \chi_3 + \chi_1\left(\frac{4 \eta}{3} - \chi_4 \right) &\ge& 0,\\
\label{condition2}
\chi_1^2\left(\frac{4 \eta}{3} - \chi_4 \right) + \lambda \chi_3(\lambda + \chi_2)
+\chi_1 \chi_2 \chi_3 &\ge& 0,\\
\label{condition3}
\lambda + \chi_1 \ge \chi_3 + \frac{4 \eta}{3} - \chi_4 &\ge& 0,
\end{eqnarray}
subject to the constraints $\lambda,\chi_1 > 0$, $\eta \ge 0$, and $\lambda \ge \eta$.  Furthermore, when
$P=0$, the bulk viscosity becomes $\zeta = \chi_3 - \chi_4$, which must be nonnegative in accordance with the second law of thermodynamics.  Hence,
we have the further requirement here that
\begin{equation}
\label{condition4}
\chi_3 \ge \chi_4.
\end{equation}

\subsection{Constant $\chi_1$, $\chi_2$, $\chi_3$, and $\chi_4$}

Let us first consider the case where 
all of $\chi$ coefficients are constant.
As noted in the previous section, there is generically
an attractor solution with $H \rightarrow H_0 = \mathrm{constant}$, with
\begin{equation}
H_0 = \frac{\chi_2 \chi_3 - \chi_1 \chi_4}{\chi_1 + \chi_3},
\end{equation}
as long as both the numerator and denominator on the right-hand side are positive.
Substituting this attractor into Eqs.\ (\ref{Fried1})-(\ref{Fried2}),
we find that $a \rho^\prime/\rho$
also evolves to a constant, given by
\begin{equation}
a \frac{\rho^\prime}{\rho} = -3 \left(\frac{\chi_2 + \chi_4}{\chi_1 + \chi_3} \right).
\label{rhoprimeeq}
\end{equation}
Furthermore, we define an effective $w_\mathrm{eff}$ given by
\begin{equation}
\label{wdef}
1 + w_\mathrm{eff} = -\frac{1}{3} a \frac{\rho^\prime}{\rho}
\end{equation}
and we note that, in particular, the special case $\chi_2 + \chi_4 = \chi_1 + \chi_3$ gives $w_\mathrm{eff} = 0$. Because the general solution of \eqref{rhoprimeeq} is 
\begin{equation}
\rho \propto a^{-3(1+w_\mathrm{eff})},
\end{equation}
one can again see the 
Cheshire Cat behavior noted earlier.  When $H \rightarrow \mathrm{constant}$, the scale factor
evolves as $a \sim e^{H_0 t}$, so the equilibrium energy density of a  fluid with constant $w_\mathrm{eff}$ will decay as a power-law
in $a$, but exponentially in $t$.

The evolution is particularly simple for the special case $\chi_3 = 0$.  In this case,
we can solve Eq.\ (\ref{Fried1cdm}) exactly to yield
\begin{equation}
\label{L-L}
H = C a^{-3/2} - \chi_4,
\end{equation}
with $C$ a constant.
As long as $\chi_4 < 0$, the Hubble parameter will evolve from initial dark matter dominated evolution
($H \sim a^{-3/2}$) to evolution resembling a cosmological constant dominated universe ($H \sim
\mathrm{constant}$).  Note, however, that $\rho$ in this case does not evolve asymptotically to a constant.
If we take $H$ to be equal to its asymptotic constant value ($H = -\chi_4$) in Eq.\
(\ref{Fried2cdm}), we obtain
\begin{equation}
\rho = \frac{3 \chi_4 (\chi_2 + \chi_4)}{D a^{3(\chi_2 + \chi_4)/\chi_1} + 1},
\end{equation}
with $D$ a constant.
The causality and stability conditions require $\chi_1 > 0$, and we need $\chi_2 + \chi_4 > 0$ to ensure a positive density.
  Although Eq.\ (\ref{Fried1cdm}) in this case
becomes exactly what one would obtain in Landau-Lifshitz theory
with constant $\zeta$ (and the evolution of $H(a)$
is therefore identical), the value of $\rho$ in the corresponding Landau-Lifshitz case evolves
to a constant.  In our approach, we once again see $\rho$ evolving to zero while viscous effects
mimic a constant-density evolution for $H$.

\subsection{Power-law density dependence of the $\chi$ parameters}

Now consider the more general case where the $\chi$ parameters are not constant but evolve as functions of $\rho$.  For simplicity, we will assume a power law behavior and take all of the
$\chi$ parameters to evolve as the same power of $\rho$, namely, $\chi_i \to \chi_i \rho^m$, $i = 1,2,3,4$,
where the $\chi_i$'s are constant. This is the analog of the case considered by \cite{LiBarrow} in Landau-Lishitz theory where it was assumed that $\zeta\sim \rho^m$.

For this case, Eqs.\ (\ref{Fried1})-(\ref{Fried2}) become
\begin{eqnarray}
\label{Fried1m}
H^\prime a + \frac{3}{2}H &=& -\frac{1}{2} \chi_3 \rho^m \frac{\rho^\prime}{\rho}
a - \frac{3}{2} \chi_4 \rho^m,\\
\label{Fried2m}
H^2 &=& \frac{1}{3} \left( \rho + \chi_1 \rho^m \frac {\rho^\prime}{\rho}aH
+ 3 \chi_2 \rho^m H \right).
\end{eqnarray}
At early times, the density of the dark energy is observed to be
negligible, and the universe is dominated
by dark matter with a density scaling as $a^{-3}$.
Thus, if we require that the ``dark energy"
corresponds to viscous corrections to the dark matter evolution, then these corrections must vanish in the limit
where $a \rightarrow 0$.  In this limit, we require $\rho$ to scale as $a^{-3}$ and $H$ to scale as $a^{-3/2}$.
Then in order for the viscous corrections to be subdominant in Eqs.\ (\ref{Fried1m})-(\ref{Fried2m})
as $a \rightarrow 0$, we need either $\chi_1 =
\chi_2$ and $\chi_3 = \chi_4$ (so that there are no
viscous corrections at all), or $m \le 1/2$.

For $0 < m \le 1/2$, we find attractor solutions with $w \rightarrow \mathrm{constant}$,
but $H \rightarrow 0$.  Hence, these do not correspond to the observed universe
if
we want viscous effects to drive the present-day accelerated expansion.
When $m < 0$, we find an attractor solution for which $H \rightarrow \infty$.
This corresponds to phantom-like behavior \cite{Caldwell:1999ew}, and can
be consistent with observations depending on the exact parameters of
the expansion.

\subsection{Inclusion of baryons}

Finally, in order to derive results that could in principle be compared with observations, we must include
both a viscous dark matter component and the nonviscous baryons (see, e.g., similar
treatments in Refs. \cite{Colistete:2007xi,LiBarrow}).  If $\rho_B$ is the density of baryons
and $\rho_D$ is the density of viscous pressureless dark matter, then we can rewrite
Eqs.\ (\ref{Fried1cdm})-(\ref{Fried2cdm}) as
\begin{eqnarray}
\label{Fried1semifinal}
H^\prime a + \frac{3}{2}H &=& -\frac{1}{2} \chi_3 \frac{\rho_D^\prime}{\rho_D} a - \frac{3}{2} \chi_4,\\
\label{Fried2semifinal}
H^2 &=& \frac{1}{3} \left( \rho_D + \rho_B + \chi_1 \frac {\rho_D^\prime}{\rho_D}aH
+ 3 \chi_2 H \right).
\end{eqnarray}
These equations can be re-expressed in the form
\begin{eqnarray}
\label{Fried2final}
H^\prime a &=& - \frac{3}{2}\left(1+ \frac{\chi_3}{\chi_1}\right)H + \frac{3}{2}\left(\frac{\chi_2 \chi_3}{\chi_1}-\chi_4\right)
+ \frac{1}{2} \left(\frac{\chi_3}{\chi_1}\right) \frac{\rho_D + \rho_B}{H},\\
\label{Fried1final}
\rho_D^\prime a &=& \frac{\rho_D}{\chi_1}\left(3H - \frac{\rho_D + \rho_B}{H} - 3 \chi_2\right).
\end{eqnarray}
The baryon density $\rho_B$ scales exactly as $a^{-3}$.  In standard nonviscous models of cold dark matter we also
have $\rho_D \propto a^{-3}$ but here the evolution of $\rho_D$ is determined instead by
Eq.\ (\ref{Fried1final}).  However, observational limits on the present-day dark matter density combined with
high-redshift estimates of $\rho_D$ from the cosmic microwave background indicate that
$\rho_D$ must evolve approximately as $a^{-3}$ up to the present.

By an appropriate choice of the $\chi$ coefficients it is possible to derive a model satisfying this constraint on $\rho_D$ that also
closely approximates $\Lambda$CDM.  We first take
\begin{eqnarray}
\label{chi3choice}
\chi_3 = - \chi_1,\\
\label{chi4choice}
\chi_4 = - \chi_2,
\end{eqnarray}
where we choose $\chi_1, \chi_2 > 0$, so that $\chi_3, \chi_4 < 0$.
This choice is consistent with the causality constraints. Indeed, if we substitute
Eqs.\ (\ref{chi3choice}) and (\ref{chi4choice}) into the
causality constraint equations (\ref{condition1})-(\ref{condition3}), we find that
there exist values for $\lambda$ and $\eta$ for which all of the constraint equations are satisfied as long as
$4/3 \ge \chi_2/\chi_1 \ge 1$.
Substituting these values for $\chi_3$ and $\chi_4$ into our evolution equations above, Eq.\ (\ref{Fried1final}) is unchanged, while
Eq.\ (\ref{Fried2final}) becomes
\begin{equation}
\label{Fried3final}
H^\prime a = - \frac{1}{2}\frac{\rho_D + \rho_B}{H}.
\end{equation}
If we neglect $\rho_B$, these equations can be solved exactly.  Dividing Eq.\ (\ref{Fried1final}) by Eq.\ (\ref{Fried3final})
yields
\begin{equation}
\frac{d\rho_D}{dH} - \frac{2}{\chi_1} \rho_D = - \frac{6}{\chi_1} H^2 + 6 \frac{\chi_2}{\chi_1}H,
\end{equation}
with solution
\begin{equation}
\rho_D = C e^{2H/\chi_1} + 3 H^2 - 3(\chi_2 - \chi_1)H - \frac{3}{2} \chi_1(\chi_2 - \chi_1),
\end{equation}
where $C$ is again a constant of integration.  At early times ($H \rightarrow \infty$) we must have $\rho = 3H^2$, so $C=0$
and our solution is
\begin{equation}
\label{Friedsolution}
\rho_D = 3 H^2 - 3(\chi_2 - \chi_1)H - \frac{3}{2} \chi_1(\chi_2 - \chi_1).
\end{equation}
Note that when $\chi_1 = \chi_2$, we once again obtain the evolution appropriate for nonviscous matter, namely
$\rho_D = 3 H^2$ (which is consistent with the fact that in this case $\zeta=0$).  Substituting Eq.\ (\ref{Friedsolution}) back into Eq.\ (\ref{Fried1final}),
we see that $\rho_D$ evolves as
\begin{equation}
\label{rhoDsolution}
\frac{\rho_D^\prime}{\rho_D}a = -3 + \frac{3}{2}(\chi_2 - \chi_1)/H.
\end{equation}

In order for Eqs.\ (\ref{Friedsolution}) and (\ref{rhoDsolution})
to mimic $\Lambda$CDM, we make one further requirement:  we take $\chi_2/\chi_1 = 1+\epsilon$,
with $0< \epsilon \ll 1 $.  (Note that this assumption means that the causality constraint $4/3 \ge \chi_2/\chi_1 \ge 1$ will
automatically be satisfied.)  With this assumption, the second term on the right-hand side of  (\ref{Friedsolution})
is always subdominant, and we have
\begin{equation}
\label{HLCDM}
H^2 = \frac{1}{3}\rho_D + \frac{1}{2} \chi_1 (\chi_2 - \chi_1). 
\end{equation}
This has the form of standard $\Lambda$CDM, where we identify $\rho_\Lambda = (3/2) \chi_1 (\chi_2 - \chi_1).$
Furthermore, Eq.\ (\ref{HLCDM}) implies that $H^2 > (1/2) \chi_1 (\chi_2 -
\chi_1)$, so that $\frac{3}{2}(\chi_2-\chi_1)/H \ll 1$.  Thus,
$\rho_D$ scales almost exactly as $a^{-3}$, as required.
Note
that this behavior for $H$, while identical
to $\Lambda$CDM, is once again an example of Cheshire Cat evolution:
the acceleration is driven by viscous effects from the
dark matter, whose equilibrium density is driven to zero by the expansion.
While we neglected $\rho_B$ in this derivation, it is easy to see
that the evolution will be unchanged when $\rho_B$ is included since
it scales in exactly the same way as $\rho_D$ ($\sim a^{-3}$), a result we have verified
with numerical integration of Eqs.\ (\ref{Fried2final})
and (\ref{Fried1final}).

Finally, we do not mean to imply that our choices for the $\chi$ coefficients in this
case are the single set of ``correct" values of these parameters for pressureless
dark matter.  Instead, we simply wish to demonstrate that causal and stable first-order viscous fluid theories
can reproduce $\Lambda$CDM for at least one choice of these parameters.  It
is quite possible that other choices for these parameters can similarly
reproduce the current observations.

%%%%%%%%%%%%%%% SECTION %%%%%%%%%%%%%%%%%%%%%%%%%%%%

\section{Conclusions}
\label{conclusions}

This work represents the first examination of cosmology with a causal, stable, first-order theory of relativistic viscous fluid dynamics (the BDNK theory).
We have shown that this effective theory approach to relativistic viscous fluids has two very attractive properties from the
standpoint of cosmology.  First, viscosity has no effect on the behavior of radiation, i.e.,
fluids with $w=1/3$.  Thus, the standard cosmology during the radiation-dominated era
simply carries over in this case without modification, including all of the successes of
BBN and the CMB.  Second, under very general conditions on the viscosity parameters,
a matter-like fluid (i.e., one with $w=0$) can generate an accelerated expansion, just
as in the case of Landau-Lifshitz theory. 

The major difference between this approach for relativistic viscous fluids and other previously-investigated models such
as the Landau-Lifshitz model
lies in the modification to Eq.\ (\ref{11bFRW}).  Previous models have modified Eq.\ (\ref{11aFRW}),
altering the effective pressure, but leaving the relationship between $H$ and $\rho$ as in
the standard cosmology (i.e., $H^2  = \rho/3$).  By altering this relationship, we decouple
the behavior of $H$ from that of the equilibrium energy density $\rho$.
Thus, while Landau-Lifshitz viscosity can cause a zero-pressure
dark matter fluid to behave as an effective
dark energy component with constant density at late times,
the BDNK theory can produce
an accelerated expansion even as the equilibrium density of the dark
matter fluid goes to zero, an effect we have dubbed the
Cheshire Cat mechanism. It is important to remark that it is $H(a)$, and not the evolution
of the dark energy density, that is actually
the observable quantity in cosmology, so the model presented here can be made
consistent with current observations of dark energy.  Indeed,
by a suitable choice of model parameters, this model can nearly
exactly mimic $\Lambda$CDM at the background level.

As we have already noted, previous attempts to use Landau-Lifshitz viscosity with
pressureless dark matter to account for the accelerated expansion
of the universe have produced acceptable results
at background level but have
foundered on the issue of perturbation growth
\cite{LiBarrow,Velten:2011bg,Velten:2012uv}.  Thus,
the results presented here are necessary but not sufficient
evidence that a viable cosmology can be constructed with
dark matter and BDNK viscous effects alone.
The next step will be to examine perturbation growth in this theory to see if it survives
this further level of scrutiny. Only then one would be able to conclusively answer whether or not viscous effects are compatible with cosmological observations.

Finally, it is important to remark that an accurate description of this vanishing equilibrium energy density behavior is, formally, beyond the regime of applicability of the hydrodynamics expansion. This occurs because in this case the out of equilibrium correction, $\mathcal{A}$, becomes larger than the equilibrium piece $\rho$. This issue is also present when considering higher-order theories, such as the generalized Israel-Stewart theories constructed using a general hydrodynamic frame in Ref.\ \cite{Noronha:2021syv}. Those should also display the properties found here, given that BDNK can be seen as the first-order truncation of such generalized 2nd order theories. Therefore, further investigation of this type of solution is needed. We defer a systematic investigation of that to future work.

\section*{Acknowledgements} 

We thank L.~Gavassino for comments on this work. Part of this work was done while F.S.B. was visiting Vanderbilt University. M.M.D. is partially supported by a Sloan Research Fellowship provided by the Alfred P. Sloan foundation, NSF grant DMS-2107701,
and a Vanderbilt's Seeding Success Grant. J.N. is partially supported by the U.S.~Department of Energy, Office of Science, Office for Nuclear Physics under Award No. DE-SC0021301. 
R.J.S. was supported in part by the Department of Energy (DE-SC0019207).

\bibliography{References}

\def\cprime{$'$}
\begin{thebibliography}{67}
\expandafter\ifx\csname natexlab\endcsname\relax\def\natexlab#1{#1}\fi
\expandafter\ifx\csname bibnamefont\endcsname\relax
  \def\bibnamefont#1{#1}\fi
\expandafter\ifx\csname bibfnamefont\endcsname\relax
  \def\bibfnamefont#1{#1}\fi
\expandafter\ifx\csname citenamefont\endcsname\relax
  \def\citenamefont#1{#1}\fi
\expandafter\ifx\csname url\endcsname\relax
  \def\url#1{\texttt{#1}}\fi
\expandafter\ifx\csname urlprefix\endcsname\relax\def\urlprefix{URL }\fi
\providecommand{\bibinfo}[2]{#2}
\providecommand{\eprint}[2][]{\url{#2}}

\bibitem[{\citenamefont{Weinberg}(1971)}]{WeinbergViscosityCosmology}
\bibinfo{author}{\bibfnamefont{S.}~\bibnamefont{Weinberg}},
  \bibinfo{journal}{Astrophys. J.} \textbf{\bibinfo{volume}{168}},
  \bibinfo{pages}{175} (\bibinfo{year}{1971}).

\bibitem[{\citenamefont{Belinsky et~al.}(1979)\citenamefont{Belinsky,
  Nikomarov, and Khalatnikov}}]{Belinsky:1979}
\bibinfo{author}{\bibfnamefont{V.~A.} \bibnamefont{Belinsky}},
  \bibinfo{author}{\bibfnamefont{E.~S.} \bibnamefont{Nikomarov}},
  \bibnamefont{and} \bibinfo{author}{\bibfnamefont{I.~M.}
  \bibnamefont{Khalatnikov}}, \bibinfo{journal}{Sov. Phys. JETP}
  \textbf{\bibinfo{volume}{50}}, \bibinfo{pages}{213} (\bibinfo{year}{1979}),
  \bibinfo{note}{[Zh. Eksp. Teor. Fiz.77,417(1979)]}.

\bibitem[{\citenamefont{Gron}(1990)}]{Gron:1990ew}
\bibinfo{author}{\bibfnamefont{O.}~\bibnamefont{Gron}},
  \bibinfo{journal}{Astrophys. Space Sci.} \textbf{\bibinfo{volume}{173}},
  \bibinfo{pages}{191} (\bibinfo{year}{1990}).

\bibitem[{\citenamefont{Murphy}(1973)}]{Murphy:1973zz}
\bibinfo{author}{\bibfnamefont{G.~L.} \bibnamefont{Murphy}},
  \bibinfo{journal}{Phys. Rev.} \textbf{\bibinfo{volume}{D8}},
  \bibinfo{pages}{4231} (\bibinfo{year}{1973}).

\bibitem[{\citenamefont{Belinsky and Khalatnikov}(1977)}]{Belinsky:1977}
\bibinfo{author}{\bibfnamefont{V.~A.} \bibnamefont{Belinsky}} \bibnamefont{and}
  \bibinfo{author}{\bibfnamefont{I.~M.} \bibnamefont{Khalatnikov}},
  \bibinfo{journal}{Sov. Phys. JETP} \textbf{\bibinfo{volume}{45}},
  \bibinfo{pages}{1} (\bibinfo{year}{1977}), \bibinfo{note}{[Zh. Eksp. Teor.
  Fiz.72,3-17(1977)]}.

\bibitem[{\citenamefont{Padmanabhan and Chitre}(1987)}]{Padmanabhan:1987dg}
\bibinfo{author}{\bibfnamefont{T.}~\bibnamefont{Padmanabhan}} \bibnamefont{and}
  \bibinfo{author}{\bibfnamefont{S.~M.} \bibnamefont{Chitre}},
  \bibinfo{journal}{Phys. Lett.} \textbf{\bibinfo{volume}{A120}},
  \bibinfo{pages}{433} (\bibinfo{year}{1987}).

\bibitem[{\citenamefont{Hiscock and Salmonson}(1991)}]{Hiscock:1991sp}
\bibinfo{author}{\bibfnamefont{W.~A.} \bibnamefont{Hiscock}} \bibnamefont{and}
  \bibinfo{author}{\bibfnamefont{J.}~\bibnamefont{Salmonson}},
  \bibinfo{journal}{Phys. Rev.} \textbf{\bibinfo{volume}{D43}},
  \bibinfo{pages}{3249} (\bibinfo{year}{1991}).

\bibitem[{\citenamefont{Pavon et~al.}(1991)\citenamefont{Pavon, Bafaluy, and
  Jou}}]{Pavon:1990qf}
\bibinfo{author}{\bibfnamefont{D.}~\bibnamefont{Pavon}},
  \bibinfo{author}{\bibfnamefont{J.}~\bibnamefont{Bafaluy}}, \bibnamefont{and}
  \bibinfo{author}{\bibfnamefont{D.}~\bibnamefont{Jou}},
  \bibinfo{journal}{Class. Quant. Grav.} \textbf{\bibinfo{volume}{8}},
  \bibinfo{pages}{347} (\bibinfo{year}{1991}).

\bibitem[{\citenamefont{Zakari and Jou}(1993)}]{Zakari:1993yhk}
\bibinfo{author}{\bibfnamefont{M.}~\bibnamefont{Zakari}} \bibnamefont{and}
  \bibinfo{author}{\bibfnamefont{D.}~\bibnamefont{Jou}},
  \bibinfo{journal}{Phys. Rev.} \textbf{\bibinfo{volume}{D48}},
  \bibinfo{pages}{1597} (\bibinfo{year}{1993}).

\bibitem[{\citenamefont{Maartens}(1995)}]{Maartens:1995wt}
\bibinfo{author}{\bibfnamefont{R.}~\bibnamefont{Maartens}},
  \bibinfo{journal}{Class. Quant. Grav.} \textbf{\bibinfo{volume}{12}},
  \bibinfo{pages}{1455} (\bibinfo{year}{1995}).

\bibitem[{\citenamefont{Zimdahl}(1996)}]{Zimdahl:1996ka}
\bibinfo{author}{\bibfnamefont{W.}~\bibnamefont{Zimdahl}},
  \bibinfo{journal}{Phys. Rev.} \textbf{\bibinfo{volume}{D53}},
  \bibinfo{pages}{5483} (\bibinfo{year}{1996}), \eprint{astro-ph/9601189}.

\bibitem[{\citenamefont{Peebles and Ratra}(2003)}]{Peebles:2002gy}
\bibinfo{author}{\bibfnamefont{P.~J.~E.} \bibnamefont{Peebles}}
  \bibnamefont{and} \bibinfo{author}{\bibfnamefont{B.}~\bibnamefont{Ratra}},
  \bibinfo{journal}{Rev. Mod. Phys.} \textbf{\bibinfo{volume}{75}},
  \bibinfo{pages}{559} (\bibinfo{year}{2003}), \eprint{astro-ph/0207347}.

\bibitem[{\citenamefont{Fabris et~al.}(2006)\citenamefont{Fabris, Goncalves,
  and de~Sa}}]{Fabris:2005ts}
\bibinfo{author}{\bibfnamefont{J.~C.} \bibnamefont{Fabris}},
  \bibinfo{author}{\bibfnamefont{S.~V.~B.} \bibnamefont{Goncalves}},
  \bibnamefont{and} \bibinfo{author}{\bibfnamefont{R.~R.} \bibnamefont{de~Sa}},
  \bibinfo{journal}{Gen. Rel. Grav.} \textbf{\bibinfo{volume}{38}},
  \bibinfo{pages}{495} (\bibinfo{year}{2006}), \eprint{astro-ph/0503362}.

\bibitem[{\citenamefont{Colistete et~al.}(2007)\citenamefont{Colistete, Fabris,
  Tossa, and Zimdahl}}]{Colistete:2007xi}
\bibinfo{author}{\bibfnamefont{R.}~\bibnamefont{Colistete}},
  \bibinfo{author}{\bibfnamefont{J.~C.} \bibnamefont{Fabris}},
  \bibinfo{author}{\bibfnamefont{J.}~\bibnamefont{Tossa}}, \bibnamefont{and}
  \bibinfo{author}{\bibfnamefont{W.}~\bibnamefont{Zimdahl}},
  \bibinfo{journal}{Phys. Rev.} \textbf{\bibinfo{volume}{D76}},
  \bibinfo{pages}{103516} (\bibinfo{year}{2007}), \eprint{0706.4086}.

\bibitem[{\citenamefont{Li and Barrow}(2009)}]{LiBarrow}
\bibinfo{author}{\bibfnamefont{B.}~\bibnamefont{Li}} \bibnamefont{and}
  \bibinfo{author}{\bibfnamefont{J.~D.} \bibnamefont{Barrow}},
  \bibinfo{journal}{Phys. Rev. D} \textbf{\bibinfo{volume}{79}},
  \bibinfo{pages}{103521} (\bibinfo{year}{2009}).

\bibitem[{\citenamefont{Avelino and Nucamendi}(2009)}]{Avelino}
\bibinfo{author}{\bibfnamefont{A.}~\bibnamefont{Avelino}} \bibnamefont{and}
  \bibinfo{author}{\bibfnamefont{U.}~\bibnamefont{Nucamendi}},
  \bibinfo{journal}{JCAP} \textbf{\bibinfo{volume}{0904}}
  (\bibinfo{year}{2009}).

\bibitem[{\citenamefont{Hipolito-Ricaldi
  et~al.}(2009)\citenamefont{Hipolito-Ricaldi, Velten, and
  Zimdahl}}]{Hipolito1}
\bibinfo{author}{\bibfnamefont{W.~S.} \bibnamefont{Hipolito-Ricaldi}},
  \bibinfo{author}{\bibfnamefont{H.}~\bibnamefont{Velten}}, \bibnamefont{and}
  \bibinfo{author}{\bibfnamefont{W.}~\bibnamefont{Zimdahl}},
  \bibinfo{journal}{JCAP} \textbf{\bibinfo{volume}{0906}}
  (\bibinfo{year}{2009}).

\bibitem[{\citenamefont{Hipolito-Ricaldi
  et~al.}(2010)\citenamefont{Hipolito-Ricaldi, Velten, and
  Zimdahl}}]{Hipolito2}
\bibinfo{author}{\bibfnamefont{W.~S.} \bibnamefont{Hipolito-Ricaldi}},
  \bibinfo{author}{\bibfnamefont{H.}~\bibnamefont{Velten}}, \bibnamefont{and}
  \bibinfo{author}{\bibfnamefont{W.}~\bibnamefont{Zimdahl}},
  \bibinfo{journal}{Phys. Rev. D} \textbf{\bibinfo{volume}{82}},
  \bibinfo{pages}{063507} (\bibinfo{year}{2010}).

\bibitem[{\citenamefont{Gagnon and Lesgourgues}(2011)}]{Gagnon}
\bibinfo{author}{\bibfnamefont{J.-S.} \bibnamefont{Gagnon}} \bibnamefont{and}
  \bibinfo{author}{\bibfnamefont{J.}~\bibnamefont{Lesgourgues}},
  \bibinfo{journal}{JCAP} \textbf{\bibinfo{volume}{1109}}
  (\bibinfo{year}{2011}).

\bibitem[{\citenamefont{Piattella et~al.}(2011)\citenamefont{Piattella, Fabris,
  and Zimdahl}}]{Piattella_et_al}
\bibinfo{author}{\bibfnamefont{O.~F.} \bibnamefont{Piattella}},
  \bibinfo{author}{\bibfnamefont{J.~C.} \bibnamefont{Fabris}},
  \bibnamefont{and} \bibinfo{author}{\bibfnamefont{W.}~\bibnamefont{Zimdahl}},
  \bibinfo{journal}{JCAP} \textbf{\bibinfo{volume}{1105}}, \bibinfo{pages}{029}
  (\bibinfo{year}{2011}).

\bibitem[{\citenamefont{Velten and Schwarz}(2012)}]{Velten:2012uv}
\bibinfo{author}{\bibfnamefont{H.}~\bibnamefont{Velten}} \bibnamefont{and}
  \bibinfo{author}{\bibfnamefont{D.}~\bibnamefont{Schwarz}},
  \bibinfo{journal}{Phys. Rev.} \textbf{\bibinfo{volume}{D86}},
  \bibinfo{pages}{083501} (\bibinfo{year}{2012}), \eprint{1206.0986}.

\bibitem[{\citenamefont{Velten and
  Schwarz}(2011{\natexlab{a}})}]{VeltenSchwarz}
\bibinfo{author}{\bibfnamefont{H.}~\bibnamefont{Velten}} \bibnamefont{and}
  \bibinfo{author}{\bibfnamefont{D.}~\bibnamefont{Schwarz}},
  \bibinfo{journal}{JCAP} \textbf{\bibinfo{volume}{1109}}, \bibinfo{pages}{16}
  (\bibinfo{year}{2011}{\natexlab{a}}).

\bibitem[{\citenamefont{Brevik and Gron}(2013)}]{BG}
\bibinfo{author}{\bibfnamefont{I.}~\bibnamefont{Brevik}} \bibnamefont{and}
  \bibinfo{author}{\bibfnamefont{O.}~\bibnamefont{Gron}}, \bibinfo{journal}{in
  Recent Advances in Cosmology, A. Travena and B. Soren (eds.) Nova Science
  Publishers}  (\bibinfo{year}{2013}).

\bibitem[{\citenamefont{Disconzi et~al.}(2015)\citenamefont{Disconzi, Kephart,
  and Scherrer}}]{Disconzi_Kephart_Scherrer_2015}
\bibinfo{author}{\bibfnamefont{M.~M.} \bibnamefont{Disconzi}},
  \bibinfo{author}{\bibfnamefont{T.~W.} \bibnamefont{Kephart}},
  \bibnamefont{and} \bibinfo{author}{\bibfnamefont{R.~J.}
  \bibnamefont{Scherrer}}, \bibinfo{journal}{Phys. Rev. D}
  \textbf{\bibinfo{volume}{91}}, \bibinfo{pages}{043532}
  (\bibinfo{year}{2015}).

\bibitem[{\citenamefont{Disconzi et~al.}(2017)\citenamefont{Disconzi, Kephart,
  and Scherrer}}]{DisconziKephartScherrerNew}
\bibinfo{author}{\bibfnamefont{M.~M.} \bibnamefont{Disconzi}},
  \bibinfo{author}{\bibfnamefont{T.~W.} \bibnamefont{Kephart}},
  \bibnamefont{and} \bibinfo{author}{\bibfnamefont{R.~J.}
  \bibnamefont{Scherrer}}, \bibinfo{journal}{International Journal of Modern
  Physics. D} \textbf{\bibinfo{volume}{26}}, \bibinfo{pages}{1750146}
  (\bibinfo{year}{2017}).

\bibitem[{\citenamefont{Cruz et~al.}(2020)\citenamefont{Cruz, Gonz\'alez, and
  Palma}}]{Cruz:2018psw}
\bibinfo{author}{\bibfnamefont{N.}~\bibnamefont{Cruz}},
  \bibinfo{author}{\bibfnamefont{E.}~\bibnamefont{Gonz\'alez}},
  \bibnamefont{and} \bibinfo{author}{\bibfnamefont{G.}~\bibnamefont{Palma}},
  \bibinfo{journal}{Gen. Rel. Grav.} \textbf{\bibinfo{volume}{52}},
  \bibinfo{pages}{62} (\bibinfo{year}{2020}), \eprint{1812.05009}.

\bibitem[{\citenamefont{Cruz et~al.}(2021)\citenamefont{Cruz, Gonz\'alez, and
  Palma}}]{Cruz:2019uya}
\bibinfo{author}{\bibfnamefont{N.}~\bibnamefont{Cruz}},
  \bibinfo{author}{\bibfnamefont{E.}~\bibnamefont{Gonz\'alez}},
  \bibnamefont{and} \bibinfo{author}{\bibfnamefont{G.}~\bibnamefont{Palma}},
  \bibinfo{journal}{Mod. Phys. Lett. A} \textbf{\bibinfo{volume}{36}},
  \bibinfo{pages}{2150032} (\bibinfo{year}{2021}), \eprint{1906.04570}.

\bibitem[{\citenamefont{Brevik et~al.}(2017)\citenamefont{Brevik, Gron,
  de~Haro, Odintsov, and Saridakis}}]{Brevik:2017msy}
\bibinfo{author}{\bibfnamefont{I.}~\bibnamefont{Brevik}},
  \bibinfo{author}{\bibfnamefont{O.}~\bibnamefont{Gron}},
  \bibinfo{author}{\bibfnamefont{J.}~\bibnamefont{de~Haro}},
  \bibinfo{author}{\bibfnamefont{S.~D.} \bibnamefont{Odintsov}},
  \bibnamefont{and} \bibinfo{author}{\bibfnamefont{E.~N.}
  \bibnamefont{Saridakis}}, \bibinfo{journal}{Int. J. Mod. Phys.}
  \textbf{\bibinfo{volume}{D26}}, \bibinfo{pages}{1730024}
  (\bibinfo{year}{2017}), \eprint{1706.02543}.

\bibitem[{\citenamefont{Eckart}(1940)}]{EckartViscous}
\bibinfo{author}{\bibfnamefont{C.}~\bibnamefont{Eckart}},
  \bibinfo{journal}{Physical Review} \textbf{\bibinfo{volume}{58}},
  \bibinfo{pages}{919} (\bibinfo{year}{1940}).

\bibitem[{\citenamefont{Landau and Lifshitz}(1987)}]{LandauLifshitzFluids}
\bibinfo{author}{\bibfnamefont{L.~D.} \bibnamefont{Landau}} \bibnamefont{and}
  \bibinfo{author}{\bibfnamefont{E.~M.} \bibnamefont{Lifshitz}},
  \emph{\bibinfo{title}{Fluid Mechanics - Volume 6 (Corse of Theoretical
  Physics)}} (\bibinfo{publisher}{Butterworth-Heinemann},
  \bibinfo{year}{1987}), \bibinfo{edition}{2nd} ed., ISBN
  \bibinfo{isbn}{0750627670}.

\bibitem[{\citenamefont{Pichon}(1965)}]{PichonViscous}
\bibinfo{author}{\bibfnamefont{G.}~\bibnamefont{Pichon}},
  \bibinfo{journal}{Annales de l'I.H.P. Physique th{\'e}orique}
  \textbf{\bibinfo{volume}{2}}, \bibinfo{pages}{21} (\bibinfo{year}{1965}).

\bibitem[{\citenamefont{Hiscock and
  Lindblom}(1985)}]{Hiscock_Lindblom_instability_1985}
\bibinfo{author}{\bibfnamefont{W.~A.} \bibnamefont{Hiscock}} \bibnamefont{and}
  \bibinfo{author}{\bibfnamefont{L.}~\bibnamefont{Lindblom}},
  \bibinfo{journal}{Phys. Rev. D} \textbf{\bibinfo{volume}{31}},
  \bibinfo{pages}{725} (\bibinfo{year}{1985}).

\bibitem[{\citenamefont{Gavassino}(2021)}]{Gavassino:2021owo}
\bibinfo{author}{\bibfnamefont{L.}~\bibnamefont{Gavassino}}
  (\bibinfo{year}{2021}), \eprint{2111.05254}.

\bibitem[{\citenamefont{Rezzolla and Zanotti}(2013)}]{Rezzolla_Zanotti_book}
\bibinfo{author}{\bibfnamefont{L.}~\bibnamefont{Rezzolla}} \bibnamefont{and}
  \bibinfo{author}{\bibfnamefont{O.}~\bibnamefont{Zanotti}},
  \emph{\bibinfo{title}{Relativistic hydrodynamics}}
  (\bibinfo{publisher}{Oxford University Press}, \bibinfo{address}{New York},
  \bibinfo{year}{2013}).

\bibitem[{\citenamefont{Israel and Stewart}(1979)}]{MIS-5}
\bibinfo{author}{\bibfnamefont{W.}~\bibnamefont{Israel}} \bibnamefont{and}
  \bibinfo{author}{\bibfnamefont{J.~M.} \bibnamefont{Stewart}},
  \bibinfo{journal}{Proc. R. Soc. London, Ser. A}
  \textbf{\bibinfo{volume}{365}}, \bibinfo{pages}{43} (\bibinfo{year}{1979}).

\bibitem[{\citenamefont{Hiscock and
  Lindblom}(1983)}]{Hiscock_Lindblom_stability_1983}
\bibinfo{author}{\bibfnamefont{W.~A.} \bibnamefont{Hiscock}} \bibnamefont{and}
  \bibinfo{author}{\bibfnamefont{L.}~\bibnamefont{Lindblom}},
  \bibinfo{journal}{Annals of Physics} \textbf{\bibinfo{volume}{151}},
  \bibinfo{pages}{466} (\bibinfo{year}{1983}).

\bibitem[{\citenamefont{Bemfica
  et~al.}(2019{\natexlab{a}})\citenamefont{Bemfica, Disconzi, and
  Noronha}}]{Bemfica:2019cop}
\bibinfo{author}{\bibfnamefont{F.~S.} \bibnamefont{Bemfica}},
  \bibinfo{author}{\bibfnamefont{M.~M.} \bibnamefont{Disconzi}},
  \bibnamefont{and} \bibinfo{author}{\bibfnamefont{J.}~\bibnamefont{Noronha}},
  \bibinfo{journal}{Phys. Rev. Lett.} \textbf{\bibinfo{volume}{122}},
  \bibinfo{pages}{221602} (\bibinfo{year}{2019}{\natexlab{a}}),
  \eprint{1901.06701}.

\bibitem[{\citenamefont{Bemfica et~al.}(2021)\citenamefont{Bemfica, Disconzi,
  Hoang, Noronha, and Radosz}}]{Bemfica:2020xym}
\bibinfo{author}{\bibfnamefont{F.~S.} \bibnamefont{Bemfica}},
  \bibinfo{author}{\bibfnamefont{M.~M.} \bibnamefont{Disconzi}},
  \bibinfo{author}{\bibfnamefont{V.}~\bibnamefont{Hoang}},
  \bibinfo{author}{\bibfnamefont{J.}~\bibnamefont{Noronha}}, \bibnamefont{and}
  \bibinfo{author}{\bibfnamefont{M.}~\bibnamefont{Radosz}},
  \bibinfo{journal}{Phys. Rev. Lett.} \textbf{\bibinfo{volume}{126}},
  \bibinfo{pages}{222301} (\bibinfo{year}{2021}), \eprint{2005.11632}.

\bibitem[{\citenamefont{Plumberg et~al.}(2021)\citenamefont{Plumberg, Almaalol,
  Dore, Noronha, and Noronha-Hostler}}]{Plumberg:2021bme}
\bibinfo{author}{\bibfnamefont{C.}~\bibnamefont{Plumberg}},
  \bibinfo{author}{\bibfnamefont{D.}~\bibnamefont{Almaalol}},
  \bibinfo{author}{\bibfnamefont{T.}~\bibnamefont{Dore}},
  \bibinfo{author}{\bibfnamefont{J.}~\bibnamefont{Noronha}}, \bibnamefont{and}
  \bibinfo{author}{\bibfnamefont{J.}~\bibnamefont{Noronha-Hostler}}
  (\bibinfo{year}{2021}), \eprint{2103.15889}.

\bibitem[{\citenamefont{Bemfica et~al.}(2018)\citenamefont{Bemfica, Disconzi,
  and Noronha}}]{BemficaDisconziNoronha}
\bibinfo{author}{\bibfnamefont{F.~S.} \bibnamefont{Bemfica}},
  \bibinfo{author}{\bibfnamefont{M.~M.} \bibnamefont{Disconzi}},
  \bibnamefont{and} \bibinfo{author}{\bibfnamefont{J.}~\bibnamefont{Noronha}},
  \bibinfo{journal}{Phys. Rev.} \textbf{\bibinfo{volume}{D98}},
  \bibinfo{pages}{104064 (26 pages)} (\bibinfo{year}{2018}),
  \eprint{1708.06255}.

\bibitem[{\citenamefont{Kovtun}(2019)}]{Kovtun:2019hdm}
\bibinfo{author}{\bibfnamefont{P.}~\bibnamefont{Kovtun}},
  \bibinfo{journal}{JHEP} \textbf{\bibinfo{volume}{10}}, \bibinfo{pages}{034}
  (\bibinfo{year}{2019}), \eprint{1907.08191}.

\bibitem[{\citenamefont{Bemfica
  et~al.}(2019{\natexlab{b}})\citenamefont{Bemfica, Disconzi, and
  Noronha}}]{Bemfica:2019knx}
\bibinfo{author}{\bibfnamefont{F.~S.} \bibnamefont{Bemfica}},
  \bibinfo{author}{\bibfnamefont{M.~M.} \bibnamefont{Disconzi}},
  \bibnamefont{and} \bibinfo{author}{\bibfnamefont{J.}~\bibnamefont{Noronha}},
  \bibinfo{journal}{Phys. Rev. D} \textbf{\bibinfo{volume}{100}},
  \bibinfo{pages}{104020} (\bibinfo{year}{2019}{\natexlab{b}}),
  \bibinfo{note}{[Erratum: Phys.Rev.D 105, 069902 (2022)]},
  \eprint{1907.12695}.

\bibitem[{\citenamefont{Bemfica et~al.}(2020)\citenamefont{Bemfica, Disconzi,
  and Noronha}}]{Bemfica:2020zjp}
\bibinfo{author}{\bibfnamefont{F.~S.} \bibnamefont{Bemfica}},
  \bibinfo{author}{\bibfnamefont{M.~M.} \bibnamefont{Disconzi}},
  \bibnamefont{and} \bibinfo{author}{\bibfnamefont{J.}~\bibnamefont{Noronha}}
  (\bibinfo{year}{2020}), \eprint{2009.11388}.

\bibitem[{\citenamefont{Hoult and Kovtun}(2020)}]{Hoult:2020eho}
\bibinfo{author}{\bibfnamefont{R.~E.} \bibnamefont{Hoult}} \bibnamefont{and}
  \bibinfo{author}{\bibfnamefont{P.}~\bibnamefont{Kovtun}},
  \bibinfo{journal}{JHEP} \textbf{\bibinfo{volume}{06}}, \bibinfo{pages}{067}
  (\bibinfo{year}{2020}), \eprint{2004.04102}.

\bibitem[{\citenamefont{Pandya and Pretorius}(2021)}]{Pandya:2021ief}
\bibinfo{author}{\bibfnamefont{A.}~\bibnamefont{Pandya}} \bibnamefont{and}
  \bibinfo{author}{\bibfnamefont{F.}~\bibnamefont{Pretorius}},
  \bibinfo{journal}{Phys. Rev. D} \textbf{\bibinfo{volume}{104}},
  \bibinfo{pages}{023015} (\bibinfo{year}{2021}), \eprint{2104.00804}.

\bibitem[{\citenamefont{Pandya et~al.}(2022{\natexlab{a}})\citenamefont{Pandya,
  Most, and Pretorius}}]{Pandya:2022pif}
\bibinfo{author}{\bibfnamefont{A.}~\bibnamefont{Pandya}},
  \bibinfo{author}{\bibfnamefont{E.~R.} \bibnamefont{Most}}, \bibnamefont{and}
  \bibinfo{author}{\bibfnamefont{F.}~\bibnamefont{Pretorius}},
  \bibinfo{journal}{Phys. Rev. D} \textbf{\bibinfo{volume}{105}},
  \bibinfo{pages}{123001} (\bibinfo{year}{2022}{\natexlab{a}}),
  \eprint{2201.12317}.

\bibitem[{\citenamefont{Bantilan et~al.}(2022)\citenamefont{Bantilan, Bea, and
  Figueras}}]{Bantilan:2022ech}
\bibinfo{author}{\bibfnamefont{H.}~\bibnamefont{Bantilan}},
  \bibinfo{author}{\bibfnamefont{Y.}~\bibnamefont{Bea}}, \bibnamefont{and}
  \bibinfo{author}{\bibfnamefont{P.}~\bibnamefont{Figueras}},
  \bibinfo{journal}{JHEP} \textbf{\bibinfo{volume}{08}}, \bibinfo{pages}{298}
  (\bibinfo{year}{2022}), \eprint{2201.13359}.

\bibitem[{\citenamefont{Pandya et~al.}(2022{\natexlab{b}})\citenamefont{Pandya,
  Most, and Pretorius}}]{Pandya:2022sff}
\bibinfo{author}{\bibfnamefont{A.}~\bibnamefont{Pandya}},
  \bibinfo{author}{\bibfnamefont{E.~R.} \bibnamefont{Most}}, \bibnamefont{and}
  \bibinfo{author}{\bibfnamefont{F.}~\bibnamefont{Pretorius}}
  (\bibinfo{year}{2022}{\natexlab{b}}), \eprint{2209.09265}.

\bibitem[{\citenamefont{Rocha et~al.}(2022)\citenamefont{Rocha, Denicol, and
  Noronha}}]{Rocha:2022ind}
\bibinfo{author}{\bibfnamefont{G.~S.} \bibnamefont{Rocha}},
  \bibinfo{author}{\bibfnamefont{G.~S.} \bibnamefont{Denicol}},
  \bibnamefont{and} \bibinfo{author}{\bibfnamefont{J.}~\bibnamefont{Noronha}},
  \bibinfo{journal}{Phys. Rev. D} \textbf{\bibinfo{volume}{106}},
  \bibinfo{pages}{036010} (\bibinfo{year}{2022}), \eprint{2205.00078}.

\bibitem[{\citenamefont{Hoult and Kovtun}(2022)}]{Hoult:2021gnb}
\bibinfo{author}{\bibfnamefont{R.~E.} \bibnamefont{Hoult}} \bibnamefont{and}
  \bibinfo{author}{\bibfnamefont{P.}~\bibnamefont{Kovtun}},
  \bibinfo{journal}{Phys. Rev. D} \textbf{\bibinfo{volume}{106}},
  \bibinfo{pages}{066023} (\bibinfo{year}{2022}), \eprint{2112.14042}.

\bibitem[{\citenamefont{Biswas et~al.}(2022)\citenamefont{Biswas, Mitra, and
  Roy}}]{Biswas:2022cla}
\bibinfo{author}{\bibfnamefont{R.}~\bibnamefont{Biswas}},
  \bibinfo{author}{\bibfnamefont{S.}~\bibnamefont{Mitra}}, \bibnamefont{and}
  \bibinfo{author}{\bibfnamefont{V.}~\bibnamefont{Roy}},
  \bibinfo{journal}{Phys. Rev. D} \textbf{\bibinfo{volume}{106}},
  \bibinfo{pages}{L011501} (\bibinfo{year}{2022}), \eprint{2202.08685}.

\bibitem[{\citenamefont{Danielsson et~al.}(2021)\citenamefont{Danielsson,
  Lehner, and Pretorius}}]{Danielsson:2021ykm}
\bibinfo{author}{\bibfnamefont{U.}~\bibnamefont{Danielsson}},
  \bibinfo{author}{\bibfnamefont{L.}~\bibnamefont{Lehner}}, \bibnamefont{and}
  \bibinfo{author}{\bibfnamefont{F.}~\bibnamefont{Pretorius}},
  \bibinfo{journal}{Phys. Rev. D} \textbf{\bibinfo{volume}{104}},
  \bibinfo{pages}{124011} (\bibinfo{year}{2021}), \eprint{2109.09814}.

\bibitem[{\citenamefont{Armas and Camilloni}(2022)}]{Armas:2022wvb}
\bibinfo{author}{\bibfnamefont{J.}~\bibnamefont{Armas}} \bibnamefont{and}
  \bibinfo{author}{\bibfnamefont{F.}~\bibnamefont{Camilloni}},
  \bibinfo{journal}{JCAP} \textbf{\bibinfo{volume}{10}}, \bibinfo{pages}{039}
  (\bibinfo{year}{2022}), \eprint{2201.06847}.

\bibitem[{\citenamefont{Weinberg}(1972)}]{Weinberg_GR_book}
\bibinfo{author}{\bibfnamefont{S.}~\bibnamefont{Weinberg}},
  \emph{\bibinfo{title}{Gravitation and Cosmology: principles and applications
  of the General Theory of Relativity}} (\bibinfo{publisher}{John Wiley \&
  Sons, Inc.}, \bibinfo{year}{1972}), \bibinfo{edition}{1st} ed., ISBN
  \bibinfo{isbn}{978-0-471-92567-5}.

\bibitem[{\citenamefont{Carroll}(1865)}]{carroll:1865}
\bibinfo{author}{\bibfnamefont{L.}~\bibnamefont{Carroll}},
  \emph{\bibinfo{title}{Alice's Adventures in Wonderland}}
  (\bibinfo{publisher}{Macmillan}, \bibinfo{year}{1865}).

\bibitem[{\citenamefont{Wald}(2010)}]{WaldBookGR1984}
\bibinfo{author}{\bibfnamefont{R.~M.} \bibnamefont{Wald}},
  \emph{\bibinfo{title}{General relativity}} (\bibinfo{publisher}{University of
  Chicago press}, \bibinfo{year}{2010}).

\bibitem[{\citenamefont{Jou et~al.}(2001)\citenamefont{Jou, Casas-Vazsquez, and
  Lebon}}]{JouetallBook}
\bibinfo{author}{\bibfnamefont{D.}~\bibnamefont{Jou}},
  \bibinfo{author}{\bibfnamefont{J.}~\bibnamefont{Casas-Vazsquez}},
  \bibnamefont{and} \bibinfo{author}{\bibfnamefont{G.}~\bibnamefont{Lebon}},
  \emph{\bibinfo{title}{Extended Irreversible Thermodynamics}}
  (\bibinfo{publisher}{Springer}, \bibinfo{year}{2001}), \bibinfo{edition}{3rd}
  ed., ISBN \bibinfo{isbn}{978-3-642-62505-3}.

\bibitem[{\citenamefont{Weinberg}(2008)}]{WeinbergCosmology}
\bibinfo{author}{\bibfnamefont{S.}~\bibnamefont{Weinberg}},
  \emph{\bibinfo{title}{Cosmology}} (\bibinfo{publisher}{Oxford University
  Press}, \bibinfo{year}{2008}), ISBN \bibinfo{isbn}{9780198526827}.

\bibitem[{\citenamefont{Cyburt et~al.}(2016)\citenamefont{Cyburt, Fields,
  Olive, and Yeh}}]{Cyburt:2015mya}
\bibinfo{author}{\bibfnamefont{R.~H.} \bibnamefont{Cyburt}},
  \bibinfo{author}{\bibfnamefont{B.~D.} \bibnamefont{Fields}},
  \bibinfo{author}{\bibfnamefont{K.~A.} \bibnamefont{Olive}}, \bibnamefont{and}
  \bibinfo{author}{\bibfnamefont{T.-H.} \bibnamefont{Yeh}},
  \bibinfo{journal}{Rev. Mod. Phys.} \textbf{\bibinfo{volume}{88}},
  \bibinfo{pages}{015004} (\bibinfo{year}{2016}), \eprint{1505.01076}.

\bibitem[{\citenamefont{Akrami and et~al.}(2020)}]{Planck2018}
\bibinfo{author}{\bibfnamefont{Y.}~\bibnamefont{Akrami}} \bibnamefont{and}
  \bibinfo{author}{\bibnamefont{et~al.}}, \bibinfo{journal}{Astron. Astrophys.}
  \textbf{\bibinfo{volume}{641}}, \bibinfo{pages}{A1} (\bibinfo{year}{2020}).

\bibitem[{\citenamefont{Bemfica
  et~al.}(2019{\natexlab{c}})\citenamefont{Bemfica, Disconzi, and
  Noronha}}]{BemficaDisconziNoronha_IS_bulk}
\bibinfo{author}{\bibfnamefont{F.~S.} \bibnamefont{Bemfica}},
  \bibinfo{author}{\bibfnamefont{M.~M.} \bibnamefont{Disconzi}},
  \bibnamefont{and} \bibinfo{author}{\bibfnamefont{J.}~\bibnamefont{Noronha}},
  \bibinfo{journal}{Physical Review Letters} \textbf{\bibinfo{volume}{122}},
  \bibinfo{pages}{221602 (11 pages)} (\bibinfo{year}{2019}{\natexlab{c}}).

\bibitem[{\citenamefont{Velten and
  Schwarz}(2011{\natexlab{b}})}]{Velten:2011bg}
\bibinfo{author}{\bibfnamefont{H.}~\bibnamefont{Velten}} \bibnamefont{and}
  \bibinfo{author}{\bibfnamefont{D.~J.} \bibnamefont{Schwarz}},
  \bibinfo{journal}{JCAP} \textbf{\bibinfo{volume}{1109}}, \bibinfo{pages}{016}
  (\bibinfo{year}{2011}{\natexlab{b}}), \eprint{1107.1143}.

\bibitem[{\citenamefont{Barbosa et~al.}(2017)\citenamefont{Barbosa, Velten,
  Fabris, and Ramos}}]{Barbosa:2017ojt}
\bibinfo{author}{\bibfnamefont{C.~M.~S.} \bibnamefont{Barbosa}},
  \bibinfo{author}{\bibfnamefont{H.}~\bibnamefont{Velten}},
  \bibinfo{author}{\bibfnamefont{J.~C.} \bibnamefont{Fabris}},
  \bibnamefont{and} \bibinfo{author}{\bibfnamefont{R.~O.} \bibnamefont{Ramos}},
  \bibinfo{journal}{Phys. Rev.} \textbf{\bibinfo{volume}{D96}},
  \bibinfo{pages}{023527} (\bibinfo{year}{2017}), \eprint{1702.07040}.

\bibitem[{\citenamefont{Aharonov et~al.}(2013)\citenamefont{Aharonov, Popescu,
  Rohrlich, and Skrzypczyk}}]{Aharonov}
\bibinfo{author}{\bibfnamefont{Y.}~\bibnamefont{Aharonov}},
  \bibinfo{author}{\bibfnamefont{S.}~\bibnamefont{Popescu}},
  \bibinfo{author}{\bibfnamefont{D.}~\bibnamefont{Rohrlich}}, \bibnamefont{and}
  \bibinfo{author}{\bibfnamefont{P.}~\bibnamefont{Skrzypczyk}},
  \bibinfo{journal}{New Journal of Physics} \textbf{\bibinfo{volume}{15}},
  \bibinfo{pages}{113015} (\bibinfo{year}{2013}).

\bibitem[{\citenamefont{Strogatz}(2000)}]{strogatz:2000}
\bibinfo{author}{\bibfnamefont{S.~H.} \bibnamefont{Strogatz}},
  \emph{\bibinfo{title}{Nonlinear Dynamics and Chaos: With Applications to
  Physics, Biology, Chemistry and Engineering}} (\bibinfo{publisher}{Westview
  Press}, \bibinfo{year}{2000}).

\bibitem[{\citenamefont{Caldwell}(2002)}]{Caldwell:1999ew}
\bibinfo{author}{\bibfnamefont{R.~R.} \bibnamefont{Caldwell}},
  \bibinfo{journal}{Phys. Lett.} \textbf{\bibinfo{volume}{B545}},
  \bibinfo{pages}{23} (\bibinfo{year}{2002}), \eprint{astro-ph/9908168}.

\bibitem[{\citenamefont{Noronha et~al.}(2022)\citenamefont{Noronha,
  Spali\'nski, and Speranza}}]{Noronha:2021syv}
\bibinfo{author}{\bibfnamefont{J.}~\bibnamefont{Noronha}},
  \bibinfo{author}{\bibfnamefont{M.}~\bibnamefont{Spali\'nski}},
  \bibnamefont{and} \bibinfo{author}{\bibfnamefont{E.}~\bibnamefont{Speranza}},
  \bibinfo{journal}{Phys. Rev. Lett.} \textbf{\bibinfo{volume}{128}},
  \bibinfo{pages}{252302} (\bibinfo{year}{2022}), \eprint{2105.01034}.

\end{thebibliography}

\end{document}